\documentclass [12pt]{article}

\usepackage{tabularx}
\usepackage[english]{babel}
\usepackage[dvips]{graphicx}
\usepackage{indentfirst}
\usepackage{epsfig}
\begin{document}

\title{New Phenomena in NC Field Theory and
Emergent Spacetime Geometry\footnote{A contribution to the Constantine workshop on Astronomy and Astrophysics, june 2010.}
}

\author{Badis Ydri\\
Institute of Physics\\
BM Annaba University,BP 12-23000-Annaba-Algeria }

\maketitle

\begin{abstract}
We give a brief review of two nonperturbative phenomena typical of noncommutative field theory which are known to lead to the perturbative instability known as the UV-IR mixing. The first phenomena concerns the emergence/evaporation of  spacetime geometry in matrix models which describe perturbative noncommutative gauge theory on fuzzy backgrounds. In particular we show that the transition from a geometrical background to a matrix phase makes the description of noncommutative gauge theory in terms of fields via the Weyl map only valid  below a critical value $g_*$. The second phenomena concerns the appearance of a nonuniform ordered phase in noncommutative scalar ${\phi}^4$ field theory and the spontaneous symmetry breaking of translational/rotational invariance which happens even in two dimensions. We argue that this phenomena also originates in the underlying matrix degrees of freedom of the noncommutative field theory. Furthermore it is conjectured that in addition to the usual WF fixed point at $\theta=0$ there must exist  a novel  fixed point at $\theta=\infty$ corresponding to the quartic hermitian matrix model. 
\end{abstract}
Classification: 11.10.Nx,11.15.Tk,11.30.Pb

\maketitle

\section{Noncommutative Geometry, String Theory and Matrix Models}
\subsection{Noncommutativity, Quantum Mechanics and General Relativity}

Spacetime noncommutativity is inspired by quantum mechanics. When the classical phase space is quantized we replace the canonical positions and momenta $x_i$ and $p_j$ by hermitian operators $\hat{x}_i$ and $\hat{p}_j$ satisfying
\begin{eqnarray}
[\hat{x}_i,\hat{p}_j]=i{\hbar}{\delta}_{ij}.
\end{eqnarray}
The quantum phase space is seen to be fuzzy, i.e points are replaced with cells due to  Heisenberg uncertainty principle
\begin{eqnarray}
{\Delta}x{\Delta}p{\geq}\frac{1}{2}\hbar.
\end{eqnarray} 
Von Neumann called this ``pointless geometry'' (see for example the introduction of \cite{Szabo:2001kg}) and the so-called Von-Neumann algebras can be viwed as marking the birth of noncommutative geometry \cite{Connes:1994yd}. The commutative limit is the quasiclassical limit $\hbar {\longrightarrow}0$.

Doplicher,Frednhagen and Roberts \cite{Doplicher:1994tu} gave arguments for the need of noncommutative structures at the Planck scale based on QM and classical general relativity. Spacetime at very large scales is a smooth manifold locally modeled on Minkowskian spacetime. Necessarily this picture breakes down at some distance scale. Measuring the coordinate $x$ of an event with an accuracy $a$ will cause an uncertaintiy in momentum of the order of $1/a$. An energy of the order $1/a$ is transmitted to the system and concentrated around $x$. This will generate a gravitational field. The smaller the uncertainty $a$ the larger the gravitational field which can then trape any signal from the event. At this stage localization loose all meaning and the manifold picture breakes down.

\subsection{Noncommutative Field Theory and Matrix Models}

Let us consider the Lagrangian 
\begin{eqnarray}
{\cal L}_m=\frac{1}{2}m(\frac{dx_i}{dt})^2-\frac{dx_i}{dt}.{A}_i~,~A_i=-\frac{B}{2}{\epsilon}_{ij}x_j.\nonumber
\end{eqnarray}
After quantization the momentum space becomes noncommutative,viz
\begin{eqnarray}
[{\pi}_i,{\pi}_j]=iB{\epsilon}_{ij}~,~{\pi}_i=m\frac{d{x}_i}{dt}.
\end{eqnarray} 
Spatial noncommutativity arises as $m{\longrightarrow}0$, i.e from
\begin{eqnarray}
{\cal L}_0=-\frac{B}{2}{\epsilon}_{ij}\frac{d{x}_i}{dt}x_j.
\end{eqnarray}
In this case we have
\begin{eqnarray}
[{x}_i,{x}_j]=i{\theta}{\epsilon}_{ij}~,~{\theta}=\frac{1}{B}.
\end{eqnarray} 
 The limit $m{\longrightarrow}0$ keeping $B$ fixed is the projection onto the lowest Landau level (recall that the mass gap is ${B}/{m}$). This projection is also achieved in the limit $B{\longrightarrow}{\infty}$ keeping $m$ fixed.

The same situation happens in string
 theory. The dynamics  of open strings moving in a flat space in the presence of a Neveu-Schwarz B-field   and with Dp-branes is equivalent to leading order in the string tension to a gauge theory on  a Moyal-Weyl space ${\bf R}^d_{\theta}$ \cite{Seiberg:1999vs}. Extension of this result to curved spaces is also possible at least in the case of open strings moving in a curved space with  ${\bf S}^3$ metric. The resulting effective gauge theory lives on a noncommutative fuzzy sphere ${\bf S}^2_N$ \cite{Alekseev:1999bs}. 

Noncommutative field theory is a field theory based on a noncommutative spacetime. The most important example is field theory on Moyal-Weyl spaces ${\bf R}^d_{\theta}$. The coordinates on ${\bf R}^d_{\theta}$  are operators which satisfy 
\begin{eqnarray}
[x_{\mu},x_{\nu}]=i{\theta}_{\mu\nu}.
\end{eqnarray}
Other examples include field theories on noncommutative tori ${\bf T}_{\theta}^d$ \cite{Ambjorn:2000cs} and field theories on fuzzy spaces \cite{O'Connor:2003aj,Balachandran:2002ig}.

The IKKT Yang-Mills matrix model in $d=10$ dimensions which is also called the  IIB matrix model is postulated to give a constructive definition of type IIB superstring theory \cite{Ishibashi:1996xs}. The IKKT model exists also in $d=4,6$ dimensions. The bosonic truncation exists in $d=3$ dimensions \cite{Krauth:1998xh}. Yang-Mills theories on  noncommutative tori can be obtained as effective field theories of the bosonic parts of the IKKT matrix models \cite{Connes:1997cr}.

Yang-Mills quantum mechanics models such as the BFSS models   in various dimensions are a non-trivial escalation over the IKKT models since they involve time. It gives in $d=10$ a constructive definition of M-theory in a falt background \cite{Banks:1996vh}. The BMN model is the unique maximally supersymmetric mass deformation  of the BFSS model in $d=10$. It gives a constructive definition of M-theory in pp-wave backgrounds \cite{Berenstein:2002jq}. The BMN model admits the fuzzy sphere \cite{Hoppe:1982} as a solution of its equations of motion. Mass deformed IKKT Yang-Mills matrix models in various dimensions admit also the fuzzy sphere as a solution.

Thus connections between noncommutative geometry and matrix models run deep. It seems to indicate that matrices are more fundamental and that noncommutativity of spacetime coordinates is just a derived property. In fact in dealing with matrix models the geometry itself which here includes spacetime geometry  and the geometry of gauge fields are also derived. 

The basic  hypothesis we start from in this talk is the following:
{``geometry, noncommutative field theory and supersymmetry should be nonperturbatively regularized with finite dimensional matrix models''}. In particular we consider that matrices are the fundamental objects and that fields,noncommutativity and geometry are derived concepts. We stress that the phenomena we will observe in the matrix models employed here as nonperturbative regularizations of noncommutative field theory are genuine effects and not  artifacts of the regularizations. The regularizations considered here are given by matrix models around fuzzy backgrounds. The fuzzy sphere in particular is of paramount importance. The noncommutative rational torus is another alternative regularization which is considered for example in  \cite{Bietenholz:2006cz} with similar consequences for the noncommutative field theory.

\section{Mass Deformed IKKT Matrix Model in $3$D}
\subsection{The Model,Ground State and The Fuzzy sphere}
We start with the IKKT matrix model in $d$ dimensions. This has ${\cal N}=1$ SUSY. This is  obtained by dimensionally reducing ${U}(N)$ SYM theory in flat $d$ dimensions to $d=0$ dimension. The dynamical variables are $d$ hermitian $N\times N$ matrices $X_1$, $X_2$,...$X_d$ together with a Majorana spinor in $d$ dimensions with action given by
\begin{eqnarray}
S=-\frac{N}{4}Tr[X_{\mu},X_{\nu}]^2+ Tr\bar{\psi}{\gamma}^{\mu}[X_{\mu},\psi].
\end{eqnarray}
The partition function exists only in $d=4,6,10$. The IKKT model in $d=3$ dimensions does not exist. In $d=3,4$ the determinant of the Dirac operator is positive definite and thus there is no sign problem \cite{Krauth:1998xh,Ambjorn:2000bf}. Although the IKKT model in $d=3$ dimensions does not exist we can still consider  its bosonic part given by the action
\begin{eqnarray}
S_0=-\frac{N}{4}Tr[X_a,X_b]^2.
\end{eqnarray}
This has no geometry since  the ground state is given by commuting matrices. Thus we consider the most general quartic polynomial matrix model with $SO(3)$ symmetry given by
\begin{eqnarray}
S&=&S_0+S_1.
\end{eqnarray}
\begin{eqnarray}
S_1&=&\frac{2iN\alpha}{3}{\epsilon}_{abc}TrX_aX_bX_c+V.
\end{eqnarray}
\begin{eqnarray}
V=N\bigg[\frac{m^2}{2c_2} Tr(X_a^2)^2-{\alpha}^2\mu Tr  (X_a^2)\bigg].\label{O3matrix}
\end{eqnarray}
The cubic (Chern-Simons) term is due to Myers effect \cite{Myers:1999ps}. This is the essential ingredient in the phenomena of condensation of geometry at low temperature \cite{Azuma:2004zq}. The potential will generically make the geometry more stable. This model with $\mu=m^2=(N^2-1)/4$ is the one considered in \cite{Steinacker:2003sd}. In this case the phenomena of condensation of geometry will not be observed in any large $N$ limit. The models considered in \cite{Steinacker:2007iy} are thought to be very different from the ones considered here.

First we consider the case $V=0$. The minimum energy configuration is

\begin{eqnarray}
X_a=\alpha L_a~,~S=-\frac{N^2-1}{48}N^2{\alpha}^4.
\end{eqnarray}
The $L_a$ are the $SU(2)$ spin $\frac{N-1}{2}$ irreducible representation, viz 
\begin{eqnarray}
[L_a,L_b]=i{\epsilon}_{abc}L_c~,~L_a^2=\frac{N^2-1}{4}.
\end{eqnarray}
Let us define the coordinates operators $x_a$ given by

\begin{eqnarray}
x_a=\frac{2}{\sqrt{N^2-1}}L_a.
\end{eqnarray}
We obtain a round sphere
\begin{eqnarray}
x_1^2+x_2^2+x_3^2=1.
\end{eqnarray}
However the coordinates are noncommuting operators, i.e
\begin{eqnarray}
[x_a,x_b]=\frac{2}{\sqrt{N^2-1}}i{\epsilon}_{abc}x_c.
\end{eqnarray}
In other words we have an uncertainty principle for spatial positions. In the limit $N\longrightarrow \infty$ we recover the commuting sphere.

The minimum energy configuration for $V=0$ and $\mu=m^2$ corresponds to $U(1)$ gauge theory on the sphere. This we explain in the next two subsections.

\subsection{Spectral Triple}

The fuzzy sphere is a quantization of the ordinary sphere in which we replace the algebra $C^{\infty}(S^2)$ by the algebra $Mat_{N}$ which acts on an $N-$dimensional Hilbert space ${H}_N$ with inner product 
\begin{eqnarray}
(f,g)=\frac{1}{N}Tr(f^{+}g)~,~f,g{\in}Mat_{N}.  
\end{eqnarray}
The fuzzy sphere is a sequence of the following triples
\begin{eqnarray}
(Mat_{N},H_N,{\Delta}_N).
\end{eqnarray}
Derivations are inner defined by the
generators of the adjoint action of $SU(2)$,i.e 

\begin{eqnarray}
{\rm Ad}L_a({\phi})\equiv[L_a,{\phi}].
\end{eqnarray}
The Laplacian is
\begin{eqnarray}
{\Delta}_N=({\rm Ad}L_a)^2=[L_a,[L_a,..]].\label{24}
\end{eqnarray}
The algebra of matrices $Mat_{N}$ decomposes under the
action of the group $SU(2)$ as
\begin{eqnarray}
\frac{N-1}{2}{\otimes}\frac{N-1}{2}=0{\oplus}1{\oplus}2{\oplus}..{\oplus}(N-1).
\end{eqnarray}
In other words the Laplacain has a
cut-off spectrum, i.e the eigenvalues are given by $k(k+1)$ where $k=0,1,...,N-1$.

\subsection{Gauge and Scalar Fluctuations}
We need to verify that the Laplacian ${\Delta}_N$ emerges from the matrix model.
 We consider fluctuation around the ground state. We introduce a $U(1)$ gauge field $A_a$ by
\begin{eqnarray}
X_a=\alpha(L_a+A_a).
\end{eqnarray}
The gauge coupling constant is defined by
\begin{eqnarray}
\frac{1}{g^2}=\tilde{\alpha}^4~,~\tilde{\alpha}=\alpha\sqrt{N}.
\end{eqnarray}
The curvature is given by
\begin{eqnarray}
F_{ab}&=&\frac{1}{{\alpha}^2}(i[X_a,X_b]+{\epsilon}_{abc}\alpha X_c)\nonumber\\
&=&i[L_a,A_b]-i[L_b,A_a]+{\epsilon}_{abc}A_c+i[A_a,A_b]\nonumber\\
&&\longrightarrow i{\cal L}_aA_b-i{\cal L}_bA_a+{\epsilon}_{abc}A_c~,~N{\longrightarrow}\infty.
\end{eqnarray}
The gauge fields $A_a$ and  covariant derivatives $X_a$ are elements of the free module ${Mat}_{N}{\otimes}{\bf C}^3$. We have $a=1,2,3$ because the differential calculus on the fuzzy sphere is $3$ dimensional. Besides the $2$ dimensional gauge field the model contains a scalar field which can be identified with the normal component of $A_a$. This is defined on the fuzzy sphere by
\begin{eqnarray}
{\Phi}=\frac{1}{2{\alpha}^2\sqrt{c_2}}(X_a^2-{\alpha}^2c_2)&=&\frac{1}{2}\big(x_aA_a+A_ax_a+\frac{A_a^2}{\sqrt{c_2}}\big){\longrightarrow}A_an_a~,~N{\longrightarrow}{\infty}.
\end{eqnarray}
The $U(1)$ gauge action on the fuzzy sphere ${\bf S}^2_N$ becomes 

\begin{eqnarray}
S&=&\frac{1}{4g^2N}TrF_{ab}^2-\frac{1}{2g^2N}{\epsilon}_{abc}Tr\left[\frac{1}{2}F_{ab}A_c-\frac{i}{6}[A_a,A_b]A_c\right]+V.
\end{eqnarray}
The potential is
\begin{eqnarray}
V=\frac{2m^2}{g^2N}Tr\Phi^2+\frac{{\rho}}{g^2N}Tr \Phi~,~\rho=(m^2-\mu)\sqrt{N^2-1}.
\end{eqnarray}
In the  commutative limit $N\longrightarrow \infty$ we obtain

\begin{eqnarray}
S&=&\frac{1}{4g^2}\int
\frac{d{\Omega}}{4{\pi}} \bigg[(F_{ab}^T)^2-4{\epsilon}_{abc}F_{ab}^Tn_c\Phi
-2[{\cal L}_a+A_a^T,\Phi]^2+4(1+2m^2){\Phi}^2+4\rho {\Phi}\bigg].\nonumber\\
\end{eqnarray}
Thus for zero gauge field we get the scalar action

\begin{eqnarray}
S&=&\frac{1}{4g^2}\int
\frac{d{\Omega}}{4{\pi}} \bigg[2\phi\Delta\phi +4(1+2m^2){\Phi}^2+4\rho {\Phi}\bigg].
\end{eqnarray}
This is a free scalar field on the sphere. The limit $m^2{\longrightarrow}\infty $ projects out the normal scalar field $\Phi$.

\section{Nonperturbative Phase Structure:The Matrix Phase and Emergent Geometry}
In this section we follow \cite{CastroVillarreal:2004vh,DelgadilloBlando:2008vi,DelgadilloBlando:2007vx} and references therein.
\subsection{Effective Potential and Phase Diagram For $\mu=m^2\longrightarrow\infty $}

The free energy $F$ is defined by
\begin{eqnarray}
&&e^{-F}=\int dX_a ~e^{-{S}[X]}.
\end{eqnarray}
In The Feynman-'t Hooft background field gauge we  fix the symmetry by imposing 
the covariant Lorentz gauge with parameter $
\frac{1}{\xi}=1+\frac{m^2}{c_2}$. In the limit $N{\longrightarrow}\infty$ keeping $\tilde{\alpha}$, $m^2$ and $\mu$ fixed with $m^2-\mu \geq 0$ we have $1)$ the path integral is dominated by the minimum of the action which is of the form $X_a=\alpha \phi L_a$ with some $\phi$ and $2)$ the one-loop becomes dominant. The free energy is therefore given
by the one-loop effective action evaluated at this minimum, viz  
\begin{eqnarray}
\frac{F}{N^2}\equiv V_{\rm eff}&=&\frac{3}{4}\log \tilde{\alpha}^4+\tilde{\alpha}^4\big[\frac{1}{8}{\phi}^4-\frac{1}{6}{\phi}^3+\frac{1}{8}m^2{\phi}^4-\frac{\mu}{4}{\phi}^2\big]+\ln
     \tilde{\alpha}{\phi}.
\end{eqnarray}
This effective potential contains a great deal of nonperturbative information about the phase structure of the model. The condition $V^{'}_{\rm eff}=0$ 
gives us extrema of the model. For large $\tilde{\alpha}$ and $m^2$ 
 it admits two positive solutions. The largest solution is the ground state. The second solution is the local maximum. As the coupling is decreased the local minimum and the local maximum merge and
the barrier disappears. This is the critical point of the model. The condition when the barrier disappears is $V_{\rm eff}^{''}=0$.  Solving the two equations  $V_{\rm eff}^{'}=V_{\rm eff}^{''}=0$ yield for $m^2\longrightarrow \infty$ the critical values

\begin{eqnarray}
{\phi}_{*}=\frac{1}{\sqrt{2}}~,~\tilde{\alpha}^4_{*}=\frac{8}{m^2}.\label{pre2}
\end{eqnarray}
This means that the phase transition is located at a smaller value of
the coupling constant $\tilde{\alpha}$ as $m$ is increased.  In other
words the region where the fuzzy sphere is stable is extended to lower
values of the coupling.  

\begin{figure}
\begin{center}
\includegraphics[width=5.0cm,angle=-90]{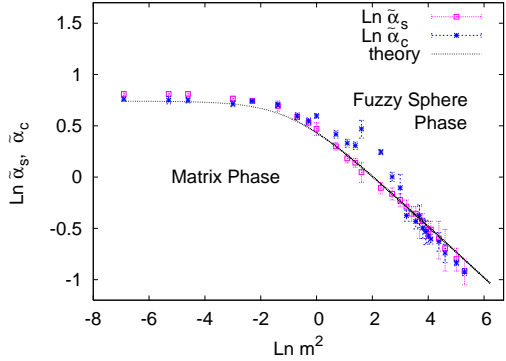}\label{phase}
\caption{The phase diagram for $\mu=m^2$ large.}
\end{center}
\end{figure}

A detailed nonperturbative Monte Carlo study of this model yields the two-dimensional phase diagram shown on figure (\ref{phase}).

\subsection{Back to Perturbation Theory: The UV-IR mixing for $V=0$}

The propagator for $V=0$ simplifies. It is given by $\frac{1}{{\Delta}_N}$. The effective action in the commutative limit is given by the expression

\begin{eqnarray}
{\Gamma}&=& \frac{1}{4g^2}\int
\frac{d{\Omega}}{4{\pi}}F_{ab}(1+2g^2{\Delta}_3)F_{ab}-\frac{1}{4g^2}{\epsilon}_{abc}\int
\frac{d{\Omega}}{4{\pi}}F_{ab}(1+2g^2{\Delta}_3)A_c+2\sqrt{N^2-1}\int\frac{d{\Omega}}{4{\pi}}\Phi \nonumber\\
&+&{\rm non~local~ quadratic ~terms}.\label{main1}
\end{eqnarray}
The $1$  in $1+2g^2{\Delta}_3$ corresponds to the classical action whereas $2g^2{\Delta}_3$ is the quantum correction. This provides a non-local renormalization of the inverse coupling constant $1/g^2$. The last terms in (\ref{main1}) are new non-local quadratic terms which have no counterpart in the classical action.

The eigenvalues of the operator ${\Delta}_3$ are given by
\begin{eqnarray}
{\Delta}_3(p)&=&\sum_{l_1,l_2}\frac{2l_1+1}{l_1(l_1+1)}\frac{2l_2+1}{l_2(l_2+1)}(1-(-1)^{l_1+l_2+p})\left\{\begin{array}{ccc}
        p & l_1 & l_2 \\
    \frac{L}{2} & \frac{L}{2} & \frac{L}{2} \end{array}\right\}^2\frac{l_2(l_2+1)}{p^2(p+1)^2}\nonumber\\
&\times &\big(l_2(l_2+1)-l_1(l_1+1)\big))\longrightarrow  -\frac{h(p)+2}{p(p+1)}~,~h(p)=-2\sum_{l=1}^{p}\frac{1}{l}.
\end{eqnarray}
In above $L+1=N$. The $1$ in $1-(-1)^{l_1+l_2+p}$ corresponds to the planar contribution whereas $(-1)^{l_1+l_2+p}$ corresponds to the non-planar contribution where $p$ is the external momentum. The fact that ${\Delta}_3\neq 0$ in the limit $N\longrightarrow 0$ means that we have a UV-IR mixing problem. We can argue that in the limit $m^2\longrightarrow \infty$ the UV-IR mixing is suppressed.	

\subsection{Thermodynamics For $V=0$: Latent Heat and Specific Heat }
\begin{figure}
\begin{center}
\includegraphics[width=5.0cm,angle=-90]{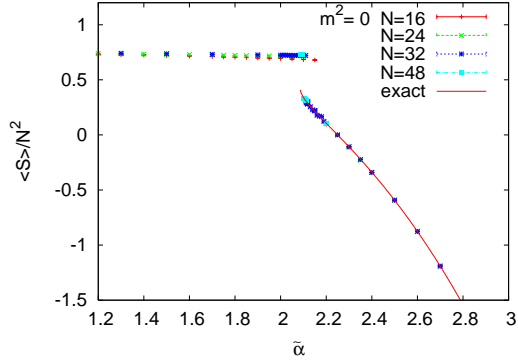}\label{U}
\caption{The observable $\frac{<S>}{N^2}$ for $m^2=0$.}
\end{center}
\end{figure}

         The inverse temperature is defined by $\beta=\tilde{\alpha}^4$. We observe in Monte Carlo simulations that the energy jumps from the value $5/12$ at low temperature to the value $3/4$ at high temperature. See figue (\ref{U}). Thus there is latent heat. This is a first order transition.The high temperature is highly interacting. Every matrix contributes $1/4$ to the energy.

\begin{figure}
\begin{center}
\includegraphics[width=5.0cm,angle=-90]{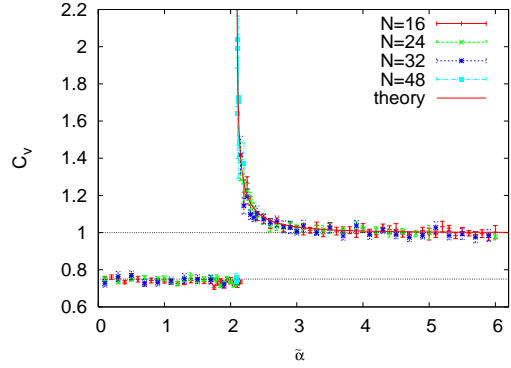}
\end{center}
\caption{The specific heat for $m^2=0$.}\label{cv}
\end{figure}

The specific heat is defined by
\begin{eqnarray}
C_v&=&<S^2>-<S>^2=-\beta\frac{d}{d\beta}\big(\frac{<S>}{\beta}\big).
\end{eqnarray}
We observe a discontinuity in the specific heat. See figure (\ref{cv}). It diverge at the transition point from the sphere side while it remains constant from the matrix side. This indicates a second order behaviour with critical fluctuations only from one side of the transition. This to our knowledge is quite novel. The critical exponent is $\alpha=1/2$, viz
\begin{eqnarray}
C_v=A_-(T-T_c)^{-\frac{1}{2}}
\end{eqnarray}
The critical value is $T_C=1/{\beta}_C=1/\tilde{\alpha}_s^4$ where $\tilde{\alpha}_s$ is given by

\begin{eqnarray}
\tilde{\alpha}_s=2.1\pm 0.1.
\end{eqnarray}
\subsection{The Order Parameter: The Radius of The Sphere}
\begin{figure}
\begin{center}
\includegraphics[width=5.0cm,angle=-90]{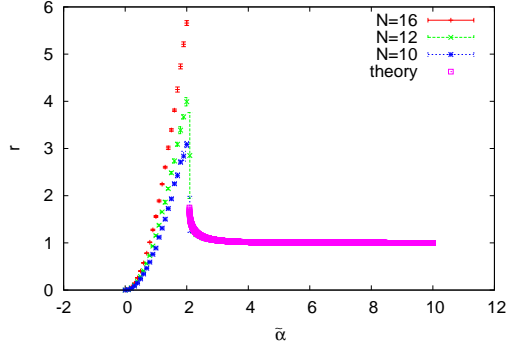}\label{radius}
\end{center}
\caption{The radius for $m^2=0$.}
\end{figure}

This is defined by
\begin{eqnarray}
\frac{1}{r}=\frac{1}{Nc_2}Tr D_a^2~,~X_a=\alpha D_a.
\end{eqnarray}
The sphere expands then evaporates as shown on figure (\ref{radius}). In other words the radius $r$ diverges at the transition point then it starts decreasing fast in the matrix phase until it reaches the value $r=0$.

The different phases of the model are characterized by 
\begin{center}
\begin{tabular}{|c|c|}
\hline
fuzzy sphere ($\tilde{\alpha}>\tilde{\alpha}_*$ )& matrix phase ($\tilde{\alpha}<\tilde{\alpha}_*$)\\
$r=1$ & $
r=0$\\
$C_v=1$  & $C_v=0.75$  \\
\hline
\end{tabular}
\end{center}
For $\mu =m^2$ the critical point is replaced by a critical line in the  $\tilde{\beta}-t$ plane where $\tilde{\beta}^4=\tilde{\alpha}^4/(1+m^2)^3$ and $t=\mu(1+m^2)$. The matrix phase persists. The nature of the transition seems to change as we increase $m^2$.

\subsection{The Matrix Phase, The $1$-Cut to N-Cut Transition and Emergent Geometry}

\begin{figure}
\begin{center}
\includegraphics[width=5.0cm,angle=-90]{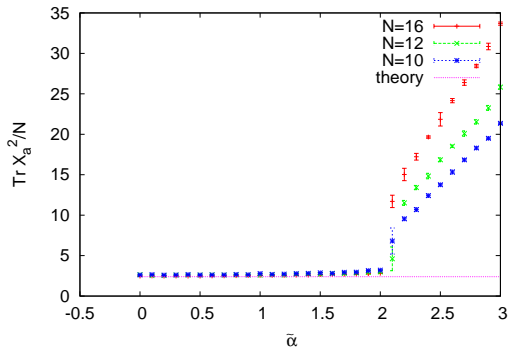}\label{radius2}
\end{center}
\caption{The radius for $m^2=0$ in the matrix phase.}
\end{figure}

The matrix phase is dominated by commuting matrices \cite{Hotta:1998en,Berenstein:2008eg}. The eigenvalues distribution of $X_3$ can be derived by assuming that the joint eigenvalues distribution of the  three commuting matrices  $X_1$, $X_2$ and $X_3$ is uniform  inside a solid ball. We obtain
\begin{eqnarray}
\rho(x)=\frac{3}{4R^3}(R^2-x^2)~,~R=2.
\end{eqnarray}
This leads to a value of the radius in the matrix phase which agrees with the exact result as shown on figure (\ref{radius2})

In the fuzzy sphere phase the matrices $X_a$  define a round sphere with a radius which scales as $N$ in the commutative limit, i.e they define a plane  whereas in the matrix phase they define a solid ball in $3$ dimensions. The scaled matrices $D_a=X_a/\alpha$  define a round sphere with finite radius in the fuzzy sphere phase whereas in the matrix phase they give a single point. We recognize two different scaling limits.

The essential ingredient in producing this transition is the Chern-Simons term in the action which is due to the Myers effect. Furthermore this transition is related to the transition found in hermitian quartic matrix models. For example the $O(3)$ matrix model given by the potential $V$ does not have any transition but when the Chern-Simons term is added to it we reproduce the one-cut to the two-cut transition. By adding the Yang-Mills term  we should then obtain a  generalization of  the one-cut to the two-cut transition. Indeed the matrix to the fuzzy sphere transition is in fact a one-cut to N-cut transition.

\begin{figure}[htbp!]
\begin{center}
\includegraphics[width=5.0cm,angle=-90]{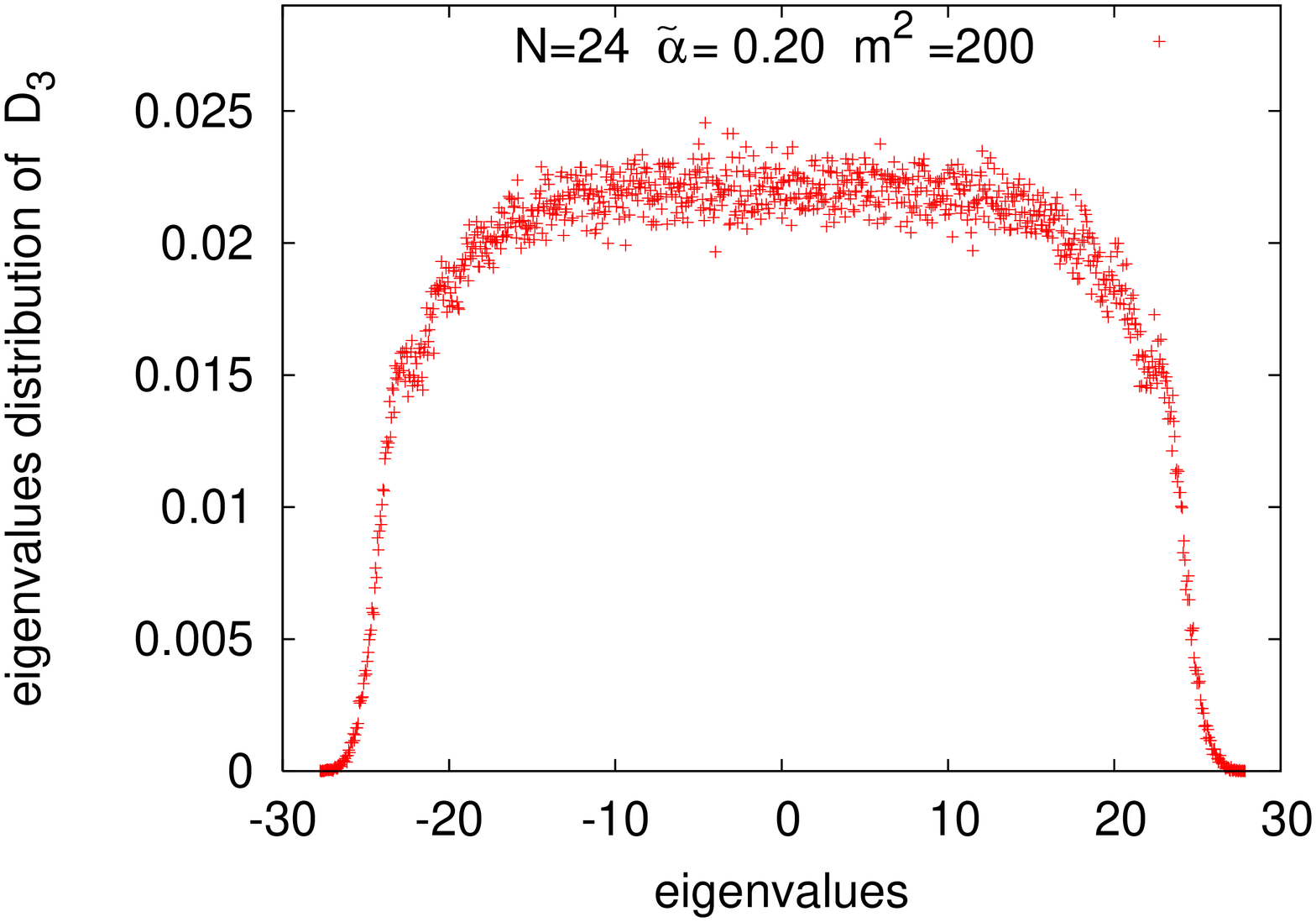}
\includegraphics[width=5.0cm,angle=-90]{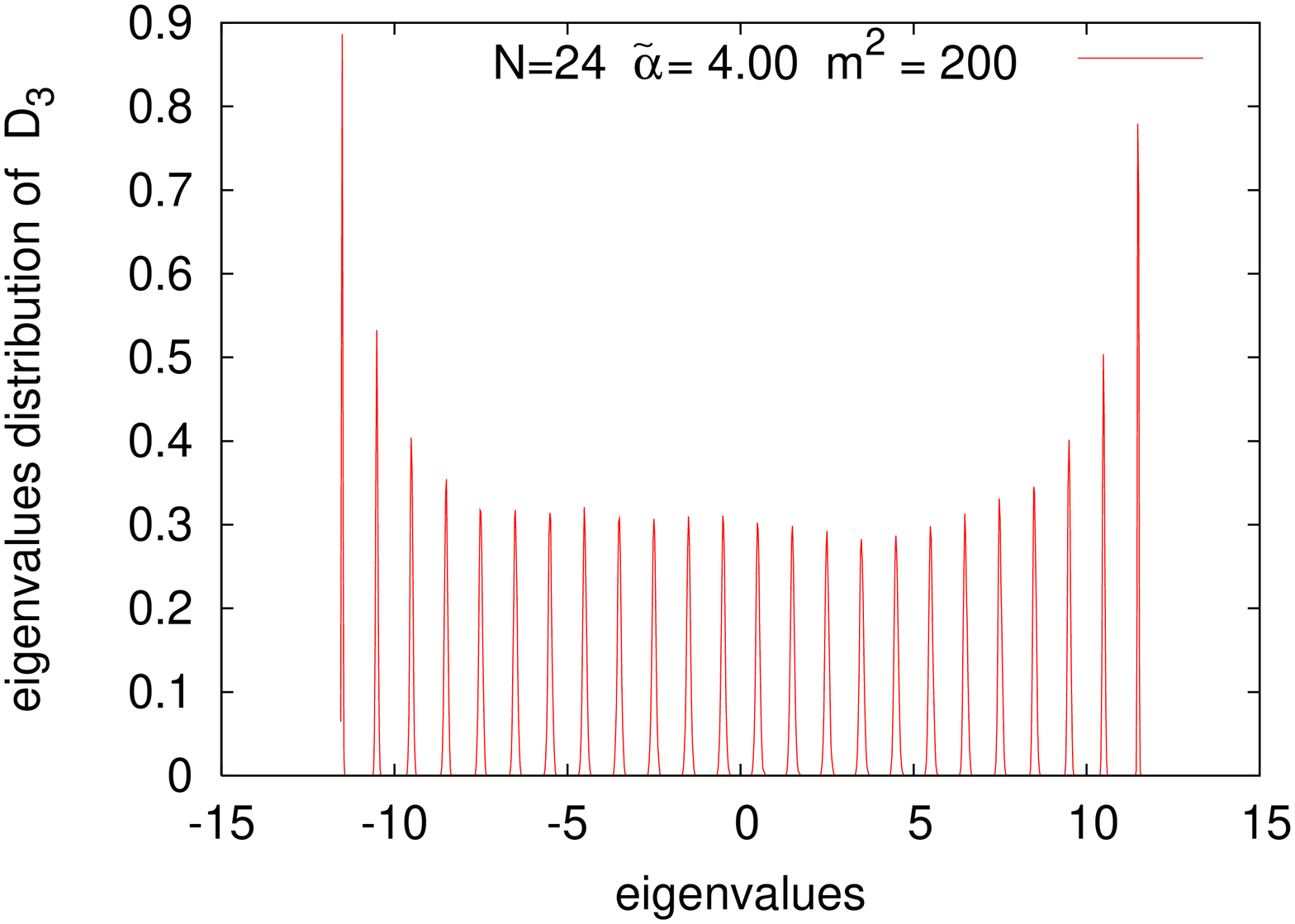}
\caption{The one-cut to N-cut transition. }\label{emerge}
\end{center}
\end{figure}

\section{Scalar Field Theory:The Non-Uniform Ordered Phase}

	\subsection{Phase Structure}

A real scalar field $\phi$ is an $N\times N$ hermitian matrix. The action is given by
\begin{eqnarray}
S=\frac{1}{N}Tr\big[\phi[L_a,[L_a,\phi]]+m^2 {\phi}^2 + \lambda {\phi}^4\big].
\end{eqnarray}
It has the correct commutative limit. Perturbatively only the tadpole diagram can diverge in the limit $N\longrightarrow \infty$. The planar and non-planar tadpole graphs are different and their difference is finite in the limit. This is the UV-IR mixing.

Non-perturbatively we find an extra phase (the non-uniform ordered phase) in which rotational invariance is spontaneously broken \cite{Martin:2004un}. The usual phases are the disordered phase ($ <Tr \phi>=0$) and the uniform ordered phase ($<Tr \phi>=\pm N\sqrt{-m^2/2\lambda}$). They are both rotationally invariant. In the  non-uniform ordered  phase 
\begin{eqnarray}
<Tr \phi >=\pm (N-2k)\sqrt{-2m^2/\lambda}
\end{eqnarray}
 where $k$ is some integer. The dominant configuration corresponds to $k=N/2$ for $N$ even and $k=(N-1)/2$ for $N$ odd. 

This phase is controlled by the quartic hermitian matrix model
\begin{eqnarray}
V=N Tr\big[m^2 {\phi}^2 + \lambda {\phi}^4\big].
\end{eqnarray}
It has a first order transition at $m_*^2=-2\sqrt{\lambda}$ from a one-cut (disordered) phase for $m^2\geq m_*^2$ to a two-cut (ordered) phase. The two-cut phase in the presence of the kinetic term becomes precisely the nonuniform ordered phase. As we will discuss shortly the kinetic term is trying to add geometry to the dynamics of the matrix $\phi$ which is at the heart of the rich phase structure we observe.

	\subsection{The Noncommutative Plane/Torus-The Stripe Phase}
The noncommutative plane should be thought of as an infinite dimensional matrix algebra not as a continuum manifold. This can be seen by introducing a periodic lattice. We obatin a noncommutative fuzzy torus where the size of the lattice is precisely the size of the matrices \cite{Ambjorn:2000cs}.  The fuzzy sphere provides also a regularization of the noncommutative plane.  Scalar field theories on these two spaces differ only in their kinetic terms. The nonuniform ordered phase is a periodically modulated phase which for small values of the coupling constant is dominated by stripes. The non-uniform ordered phase on the fuzzy sphere becomes therefore a stripe phase on the noncommutative plane/torus \cite{Ambjorn:2002nj,Bietenholz:2004xs}.

The nonuniform phase is the analogue of the matrix phase in pure gauge models in the sense that in this phase the spacetime metric is modified by quantum fluctuations of the noncommutative field theory since the Laplacian is found to be  $({\partial}_{\mu}^2)^2$ and not  ${\partial}_{\mu}^2$ \cite{Chen:2001an}. The disordered phase, the uniform ordered phase and the nonuniform ordered phase meet at a triple point possibly a Lifshitz point \cite{Gubser:2000cd,Chen:2001an}. Since the UV-IR mixing is equivalent to the fact that the two-point function goes to $\infty$ for small momenta we can  immediately conclude that the two-point function has a minimum away from zero. This is the underlying reason for the condensation of non-zero modes and as a consequence the spontaneous symmetry breaking of translational invariance and appearance of stripes \cite{Gubser:2000cd}.

We conjecture 
that there must exist two fixed points in this theory, the usual Wilson-Fisher fixed point at $\theta=0$ and a novel fixed point at $\theta=\infty$ which is intimately related to the dominance of the quartic hermitian matrix model $V$.  Indeed in the limit $\theta\longrightarrow\infty$ we obtain the planar theory (only planar graphs survive) \cite{Filk:1996dm} which is intimately related to large $N$ limits of hermitian matrix model.

 The kinetic term is trying to add a geometry to the dynamics of the matrix $\phi$ which is at the heart of the rich phase structure we observe. For small $\lambda$ the usual Ising model transition is expected and the $\theta=0$ fixed point should control the physics. We expect that the  $\theta=\infty$ fixed point should control most of the phase diagram since generically the  ${\phi}^4$ interaction is not weak. In the nonuniform ordered phase the kinetic term is very small compared to $V$ but not zero. The matrix regularization of noncommutative ${\phi}^4$ given in \cite{Grosse:2003aj} with $\Omega=1$ is a matrix model closely related to $V$ but in which the kinetic term is not zero.  It is natural to expect that this action is precisely the fixed point action corresponding to $\theta=\infty$.

\section{Summary And Outlook}

We find for $d=3$ mass deformed IKKT matrix models with global $SO(3)$ symmetry  in the limit of small deformation a  line of discontinuous transitions with a jump in the energy characteristic of a first order
transition but with divergent critical fluctuations and a divergent
specific heat which is characteristic of a second order
transition. The low temperature
phase (small values of the gauge coupling constant) is a geometrical one with gauge fields fluctuating on a round sphere. 
As the temperature increased the sphere evaporates 
in a transition to a pure matrix phase with no background geometrical
structure.  

In the limit of large deformation the transition seems to be different. Also within the fuzzy sphere phase there are strong indications for the existence of other phases which can be characterized as field theory phases.

The most important next step is to consider mass deformed IKKT model in $d=4$ dimensions. This model is well defined with supersymmetry. This will allow us to study the impact of supersymmetry on emergent geometry and vice versa. This will also be a concrete example in which the Monte Carlo method can be applied to study exact supersymmetry via matrix models.

The second important direction is to find a matrix model in which we have emergent $4$ dimensional geometry and gauge theory. We claim that fuzzy ${\bf S}^2\times {\bf S}^2$ is the correct choice.  

Also computing the phase diagram of noncommutative scalar field theory using the renormalization group method remains a very challenging task even in two dimensions. More importantly is the determination of the structure of the fixed points in this class of theories.


\begin{thebibliography}{10}

\bibitem{Doplicher:1994tu}
  S.~Doplicher, K.~Fredenhagen and J.~E.~Roberts,
  ``The Quantum structure of space-time at the Planck scale and quantum
  fields,''
  Commun.\ Math.\ Phys.\  {\bf 172}, 187 (1995)
  [arXiv:hep-th/0303037];
  ``Space-time quantization induced by classical gravity,''
  Phys.\ Lett.\  B {\bf 331}, 39 (1994).









\bibitem{Connes:1994yd}
  A.~Connes,
  ``Noncommutative geometry,''
   Academic Press,London, 1994.








\bibitem{Szabo:2001kg}
  R.~J.~Szabo,
  ``Quantum Field Theory on Noncommutative Spaces,''
  Phys.\ Rept.\  {\bf 378}, 207 (2003)
  [arXiv:hep-th/0109162].




\bibitem{Seiberg:1999vs}
  N.~Seiberg and E.~Witten,
  ``String theory and noncommutative geometry,''
  JHEP {\bf 9909}, 032 (1999)
  [arXiv:hep-th/9908142].



\bibitem{Connes:1997cr}
  A.~Connes, M.~R.~Douglas and A.~S.~Schwarz,
  ``Noncommutative geometry and matrix theory: Compactification on tori,''
  JHEP {\bf 9802}, 003 (1998)
  [arXiv:hep-th/9711162].




\bibitem{Alekseev:1999bs}
  A.~Y.~Alekseev, A.~Recknagel and V.~Schomerus,
  ``Non-commutative world-volume geometries: Branes on SU(2) and fuzzy
  spheres,''
  JHEP {\bf 9909}, 023 (1999)
  [arXiv:hep-th/9908040];
  ``Brane dynamics in background fluxes and non-commutative geometry,''
  JHEP {\bf 0005}, 010 (2000)
  [arXiv:hep-th/0003187].

\bibitem{Hoppe:1982}
  J.~Hoppe,
  ``Quantum theory of a massless relativistic surface and a two-dimensional bound state problem,''
  Ph.D thesis,MIT,1982. 
  J.~Madore,
  ``The fuzzy sphere,''
  Class.\ Quant.\ Grav.\  {\bf 9}, 69 (1992).




\bibitem{Myers:1999ps}
  R.~C.~Myers,
  ``Dielectric-branes,''
  JHEP {\bf 9912}, 022 (1999)
  [arXiv:hep-th/9910053].








\bibitem{Azuma:2004zq}
  T.~Azuma, S.~Bal, K.~Nagao and J.~Nishimura,
  ``Nonperturbative studies of fuzzy spheres in a matrix model with the
  Chern-Simons term,''
  JHEP {\bf 0405}, 005 (2004)
  [arXiv:hep-th/0401038];
  ``Perturbative dynamics of fuzzy spheres at large N,''
  JHEP {\bf 0506}, 081 (2005)
  [arXiv:hep-th/0410263].






\bibitem{CastroVillarreal:2004vh}
  P.~Castro-Villarreal, R.~Delgadillo-Blando and B.~Ydri,
  ``A gauge-invariant UV-IR mixing and the corresponding phase transition  for
  U(1) fields on the fuzzy sphere,''
  Nucl.\ Phys.\  B {\bf 704}, 111 (2005)
  [arXiv:hep-th/0405201].



\bibitem{DelgadilloBlando:2008vi}
  R.~Delgadillo-Blando, D.~O'Connor and B.~Ydri,
  ``Matrix Models, Gauge Theory and Emergent Geometry,''
  arXiv:0806.0558 [hep-th].

\bibitem{DelgadilloBlando:2007vx}
  R.~Delgadillo-Blando, D.~O'Connor and B.~Ydri,
  ``Geometry in transition: A model of emergent geometry,''
  Phys.\ Rev.\ Lett.\  {\bf 100}, 201601 (2008)
  [arXiv:0712.3011 [hep-th]].















\bibitem{Steinacker:2003sd}
  H.~Steinacker,
  ``Quantized gauge theory on the fuzzy sphere as random matrix model,''
  Nucl.\ Phys.\  B {\bf 679}, 66 (2004)
  [arXiv:hep-th/0307075].

\bibitem{Steinacker:2007iy}
  H.~Steinacker and R.~J.~Szabo,
  ``Nonabelian localization for gauge theory on the fuzzy sphere,''
  J.\ Phys.\ Conf.\ Ser.\  {\bf 103}, 012017 (2008)
  [arXiv:0708.4365 [hep-th]];
  ``Localization for Yang-Mills Theory on the Fuzzy Sphere,''
  Commun.\ Math.\ Phys.\  {\bf 278}, 193 (2008)
  [arXiv:hep-th/0701041].










\bibitem{Krauth:1998xh}
  W.~Krauth, H.~Nicolai and M.~Staudacher,
  ``Monte Carlo approach to M-theory,''
  Phys.\ Lett.\  B {\bf 431}, 31 (1998)
  [arXiv:hep-th/9803117].
  W.~Krauth and M.~Staudacher,
  ``Finite Yang-Mills integrals,''
  Phys.\ Lett.\  B {\bf 435}, 350 (1998)
  [arXiv:hep-th/9804199].
  P.~Austing and J.~F.~Wheater,
  ``Convergent Yang-Mills matrix theories,''
  JHEP {\bf 0104}, 019 (2001)
  [arXiv:hep-th/0103159].
  P.~Austing,
  ``Yang-Mills matrix theory,''
  arXiv:hep-th/0108128.

\bibitem{Ishibashi:1996xs}
  N.~Ishibashi, H.~Kawai, Y.~Kitazawa and A.~Tsuchiya,
  ``A large-N reduced model as superstring,''
  Nucl.\ Phys.\  B {\bf 498}, 467 (1997)
  [arXiv:hep-th/9612115].


\bibitem{Banks:1996vh}
  T.~Banks, W.~Fischler, S.~H.~Shenker and L.~Susskind,
  ``M theory as a matrix model: A conjecture,''
  Phys.\ Rev.\  D {\bf 55}, 5112 (1997)
  [arXiv:hep-th/9610043].


\bibitem{Berenstein:2002jq}
  D.~E.~Berenstein, J.~M.~Maldacena and H.~S.~Nastase,
  ``Strings in flat space and pp waves from N = 4 super Yang Mills,''
  JHEP {\bf 0204}, 013 (2002)
  [arXiv:hep-th/0202021].














\bibitem{Grosse:2003aj}
  H.~Grosse and R.~Wulkenhaar,
  ``Power-counting theorem for non-local matrix models and renormalisation,''
  Commun.\ Math.\ Phys.\  {\bf 254}, 91 (2005)
  [arXiv:hep-th/0305066];
  ``Renormalisation of phi**4 theory on noncommutative R**4 in the matrix
  base,''
  Commun.\ Math.\ Phys.\  {\bf 256}, 305 (2005)
  [arXiv:hep-th/0401128],
  ``Renormalisation of phi**4 theory on noncommutative R**2 in the matrix
  base,''
  JHEP {\bf 0312}, 019 (2003)
  [arXiv:hep-th/0307017].

\bibitem{Hotta:1998en}
  T.~Hotta, J.~Nishimura and A.~Tsuchiya,
  ``Dynamical aspects of large N reduced models,''
  Nucl.\ Phys.\  B {\bf 545}, 543 (1999)
  [arXiv:hep-th/9811220].

\bibitem{Berenstein:2008eg}
  D.~E.~Berenstein, M.~Hanada and S.~A.~Hartnoll,
  ``Multi-matrix models and emergent geometry,''
  JHEP {\bf 0902}, 010 (2009)
  [arXiv:0805.4658 [hep-th]].


\bibitem{Martin:2004un}
  X.~Martin,
  ``A matrix phase for the phi**4 scalar field on the fuzzy sphere,''
  JHEP {\bf 0404}, 077 (2004)
  [arXiv:hep-th/0402230].
  F.~Garcia Flores, D.~O'Connor and X.~Martin,
  ``Simulating the scalar field on the fuzzy sphere,''
  PoS {\bf LAT2005}, 262 (2006)
  [arXiv:hep-lat/0601012].
  M.~Panero,
  JHEP {\bf 0705}, 082 (2007)
  [arXiv:hep-th/0608202].



\bibitem{Ambjorn:2002nj}
  J.~Ambjorn and S.~Catterall,
  ``Stripes from (noncommutative) stars,''
  Phys.\ Lett.\  B {\bf 549}, 253 (2002)
  [arXiv:hep-lat/0209106].

\bibitem{Bietenholz:2004xs}
  W.~Bietenholz, F.~Hofheinz and J.~Nishimura,
  ``Phase diagram and dispersion relation of the non-commutative lambda  phi**4
  model in d = 3,''
  JHEP {\bf 0406}, 042 (2004)
  [arXiv:hep-th/0404020].


\bibitem{Gubser:2000cd}
  S.~S.~Gubser and S.~L.~Sondhi,
  ``Phase structure of non-commutative scalar field theories,''
  Nucl.\ Phys.\  B {\bf 605}, 395 (2001)
  [arXiv:hep-th/0006119].

\bibitem{Chen:2001an}
  G.~H.~Chen and Y.~S.~Wu,
  ``Renormalization group equations and the Lifshitz point in  noncommutative
  Landau-Ginsburg theory,''
  Nucl.\ Phys.\  B {\bf 622}, 189 (2002)
  [arXiv:hep-th/0110134].

\bibitem{Ambjorn:2000cs}
  J.~Ambjorn, Y.~M.~Makeenko, J.~Nishimura and R.~J.~Szabo,
  ``Lattice gauge fields and discrete noncommutative Yang-Mills theory,''
  JHEP {\bf 0005}, 023 (2000)
  [arXiv:hep-th/0004147],
  ``Finite N matrix models of noncommutative gauge theory,''
  JHEP {\bf 9911}, 029 (1999)
  [arXiv:hep-th/9911041].


\bibitem{Filk:1996dm}
  T.~Filk,
  ``Divergencies in a field theory on quantum space,''
  Phys.\ Lett.\  B {\bf 376}, 53 (1996).


\bibitem{Bietenholz:2006cz}
  W.~Bietenholz, J.~Nishimura, Y.~Susaki and J.~Volkholz,
  ``A non-perturbative study of 4d U(1) non-commutative gauge theory: The fate
  of one-loop instability,''
  JHEP {\bf 0610}, 042 (2006)
  [arXiv:hep-th/0608072].
  W.~Bietenholz, F.~Hofheinz and J.~Nishimura,
  ``The renormalizability of 2D Yang-Mills theory on a non-commutative
  geometry,''
  JHEP {\bf 0209}, 009 (2002)
  [arXiv:hep-th/0203151].

\bibitem{O'Connor:2003aj}
  D.~O'Connor,
  ``Field theory on low dimensional fuzzy spaces,''
  Mod.\ Phys.\ Lett.\  A {\bf 18}, 2423 (2003).

\bibitem{Balachandran:2002ig}
  A.~P.~Balachandran,
  ``Quantum spacetimes in the year 1,''
  Pramana {\bf 59}, 359 (2002)
  [arXiv:hep-th/0203259].

\bibitem{Ambjorn:2000bf}
  J.~Ambjorn, K.~N.~Anagnostopoulos, W.~Bietenholz, T.~Hotta and J.~Nishimura,
  ``Large N dynamics of dimensionally reduced $4$D $SU(N)$ super Yang-Mills
  theory,''
  JHEP {\bf 0007}, 013 (2000)
  [arXiv:hep-th/0003208].

\end{thebibliography}
\end{document}